\documentclass[aps,prl,twocolumn,showpacs,preprintnumbers,superscriptaddress,amsmath]{revtex4}
\usepackage{graphicx}
\usepackage{dcolumn}
\usepackage{bm}

\begin{document}

\title{Observation of the Preservation of the Two-Photon Coherence in Plasmon-assisted Transmission}
\author{Guo-Ping Guo\footnote[1]{gpguo@ustc.edu.cn}}
\address{Key Laboratory of Quantum Information and Department
of Physics, University of Science and Technology of China, Hefei
230026, People's Republic of China}
\author{Xi-Feng Ren}
\address{Key Laboratory of Quantum Information and Department
of Physics, University of Science and Technology of China, Hefei
230026, People's Republic of China}
\author{Yun-Feng Huang}
\address{Key Laboratory of Quantum Information and Department
of Physics, University of Science and Technology of China, Hefei
230026, People's Republic of China}
\author{Chuan-Feng Li}
\address{Key Laboratory of Quantum Information and Department
of Physics, University of Science and Technology of China, Hefei
230026, People's Republic of China}
\author{Zhe-Yu Ou}
\address{Key Laboratory of Quantum Information and Department
of Physics, University of Science and Technology of China, Hefei
230026, People's Republic of China}
\address{Department of Physics, Indiana University-Purdue University Indianapolis 402
N Blackford Street, Indianapolis, In 46202, USA}
\author{Guang-Can Guo}
\address{Key Laboratory of Quantum Information and Department
of Physics, University of Science and Technology of China, Hefei
230026, People's Republic of China}

\begin{abstract}
We experimentally study two-photon coherence in  plasmon-assisted
transmission with a two-photon Mach-Zehnder (MZ) interferometer.
Two collinear photons of identical or orthogonal polarization are
simultaneously incident on one optically thick metal film,
perforated with a periodic array of subwavelength holes. The de
Broglie wavelength of plasmon-assisted transmitted photons is
measured, which shows that two photons are re-eradiated by the
plasmons and the quantum coherence of biphoton is preserved in the
conversion process of transforming biphoton to plasmons and then
back to biphoton.
\end{abstract}
\maketitle

It has long been observed that there is an unusually high optical
transmission efficiency in metal films perforated with a periodic
array of subwavelength apertures\cite{1}. Generally, it is
believed that metal surface plays a crucial role and the
phenomenon is mediated by surface plasmons (SPs) and there is a
process of transform photon to surface plasmon and back to
photon\cite{4,5,crucial,ebbesen5}. In 2002, Ebbesen \textit{et
al.} \cite{ebbesen} first addressed the question of whether the
entanglement survives in this extraordinary enhancement light
transmission. They showed that quantum entanglement of transmitted
photon pair can be preserved when they respectively travel through
a hole array. Therefore, the macroscopic surface plasmon
polarizations, a collective excitation wave involving typically
$10^{10}$ free electrons propagating at the surface of conducting
matter, have a true quantum nature. Although several theory models
\cite{ebbesen,theory,theory2} have been proposed to explain this
quantum coherence preservation, and extraordinary enhancement
light transmission through subwavelength apertures in metal
plates, up to our knowledge, a prevalent theory has not yet been
developed. Many more experiments were preformed for a rounded
understanding of this problem\cite{a1,a2,a3,a4,a5,a6}. For
example, Young's double-slit experiment is revisited by an
experimental and theoretical study of the optical transmission of
a thin metal screen perforated by two subwavelength slits,
separated by many optical wavelengths\cite{young}. And there is
also a demonstration showing the preservation of the energy-time
entanglement of a pair of photons after a photon-plasmon-photon
conversion \cite{energy}. These works combines of two fields of
research, namely quantum information and nano-structured metal
optics. Apart from its fundamental interest, this offers a
possibility to transfer entanglement between photons and
condensed-matter, for storing, modulation and even engineering of
quantum information with subwavelength metal optics.

\begin{figure}[b]
\includegraphics[width=7.0cm]{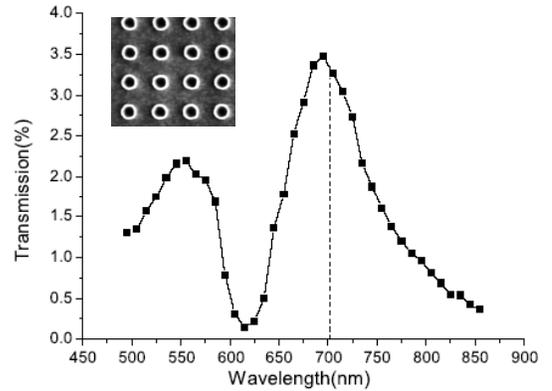}
\caption{Hole array transmittance as a function of wavelength. The dashed
vertical line indicates the wavelength of 702nm used in the experiment.
Inset, scanning electron microscope picture of part of a typical hole array.
After subsequently evaporating a 3-nm titanium bonding layer and a 135-nm
gold layer onto a 0.5-mm-thick silica glass substrate, a Focused Ion Beam
Etching system (FIB, DB235 of FEB Co.) is used to produce cylindrical holes
(200 nm diameter) arranged as a square lattice (period 600 nm). The total
area of the hole array is $30\mu m\times 30\mu m$ and it is actually made up
with four hole arrays of $15\mu m\times 15\mu m$ area for the technical
reason.}
\end{figure}

\begin{figure}[b]
\includegraphics[width=8.0cm]{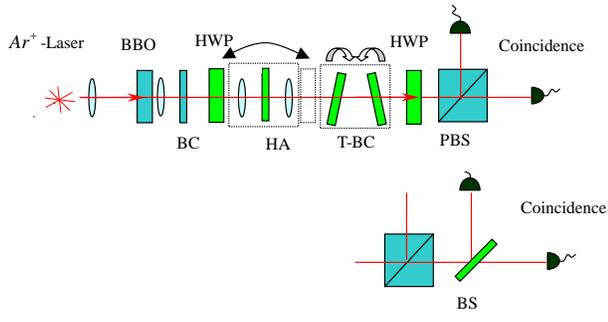}
\caption{Experimental set-up. We pump a type-II 2.0mm thick BBO
crystal with an Ar$^{+}$-Laser at 351nm. The optical pump power is
approximately 400 mW and the pump laser is vertically polarized.
Via spontaneous parametric down-conversion (SPDC) pairs of
photons(702nm) in HV polarization are produced. Reflector (351 nm)
and Interference filter IF (4-nm bandwidth centred at 702nm) are
used to reflect the pump light (not shown in the figure). The
effect of the pump photon on the background can be neglected.
Birefringent crystals(BC) are used to make the photon pairs arrive
at the hole array(HA) in the same time. A HWP(at 702nm) is placed
first before and then after the twin-lenses. We control the phase
difference in the interferometer by rotating a twin-birefringent
crystal(T-BC). Coincidences in the HV basis and HH
basis(separately shown) are recorded separately.}
\end{figure}

Since there is a process of transforming photon to surface plasmon
and then back to photon\cite{4,5,crucial,ebbesen5}, do two
simultaneously input collinear photons excite two plasmon modes or
only one? Whether or how those plasmon modes interact with each
other? How many photons will be emitted and what is the
relationship between those photons? With these problems, we
simultaneously input {\it two} collinear photons of identical or
orthogonal polarization into one optically thick metal film,
perforated with a periodic array of subwavelength holes. Previous
studies on the revealed quantum coherence preservation for single
photon, whereas our study emphasizes on two-photon quantum
coherence. It is observed that two photons are still re-eradiated
by the surface-plasmon waves tunnelling through the holes after
excited by two incident photons. We employ a Mach-Zehnder (MZ)
type two-photon polarization interferometer and measure the de
Broglie wavelength of plasmon-assisted transmitted photons. The
measured de Broglie wavelength of re-eradiated photons equals that
of Fock state $\left| 2\right\rangle $, which shows that there are
two photons re-eradiated and the quantum coherence of biphoton is
preserved in the conversion process of transforming biphoton to
plasmon and then back to biphoton.

Fig. 1 shows the hole array transmittance as a function of
wavelength. The dashed vertical line indicates the wavelength of
702 nm used in our interference experiment. The transmission of
the array at 702 nm is about 3.2\%, which is much larger than the
value of 0.55\% obtained from classical theory\cite {ebbesen8}.
After subsequently evaporating a 3-nm titanium bonding layer and a
135-nm gold layer onto a 0.5-mm-thick silica glass substrate, a
Focused Ion Beam Etching system (FIB, DB235 of FEB Co.) is used to
produce cylindrical holes (200 nm diameter) arranged as a square
lattice (period 600 nm). The total area of the hole array is
$30\mu m\times 30\mu m$ and is actually made up of four hole
arrays of $15\mu m\times 15\mu m$ area for the technical reason.
From a calculation based on the geometry of the array and the
optical constants of gold and glass, 702 nm wavelength light is
associated with the surface plasmon modes (0,$\pm $1) or ($\pm
$1,0) on the glass-metal interface. The metal plate is set between
two lenses of 35 cm focal length, so that the light is normally
incident on the hole array, where it has a cross section diameter
of about $20\mu m$ and covers hundreds of holes.

The standard method of type-II collinear spontaneous parametric
down-conversion is employed to generate a pair of degenerate
photons in 702 nm wavelength, as shown in Fig. 2. A 400 mW
continuous-wave Ar$^{+}$-Laser at the wavelength of 351 nm is
directed onto a type-II 2.0mm thick BBO crystal. The pump laser is
vertically polarized and the down-converted photon pair of 702 nm
wavelength are in $\left| HV\right\rangle $ mode propagating in
the same direction. A reflector (351 nm) and an interference
filter IF (4-nm bandwidth centered at 702nm) are employed to
reflect and filter the pump light (not shown in the figure). A
birefringent crystal (BC) is used to make the two collinear
down-converted photon pair in the $\left| HV\right\rangle $ mode
simultaneously arrive at the half wavelength plate (HWP) or the
hole array. Silicon avalanche photodiode (APD) photon counters are
used to record counts.

\begin{figure}[b]
\includegraphics[width=4.0cm,height=5cm]{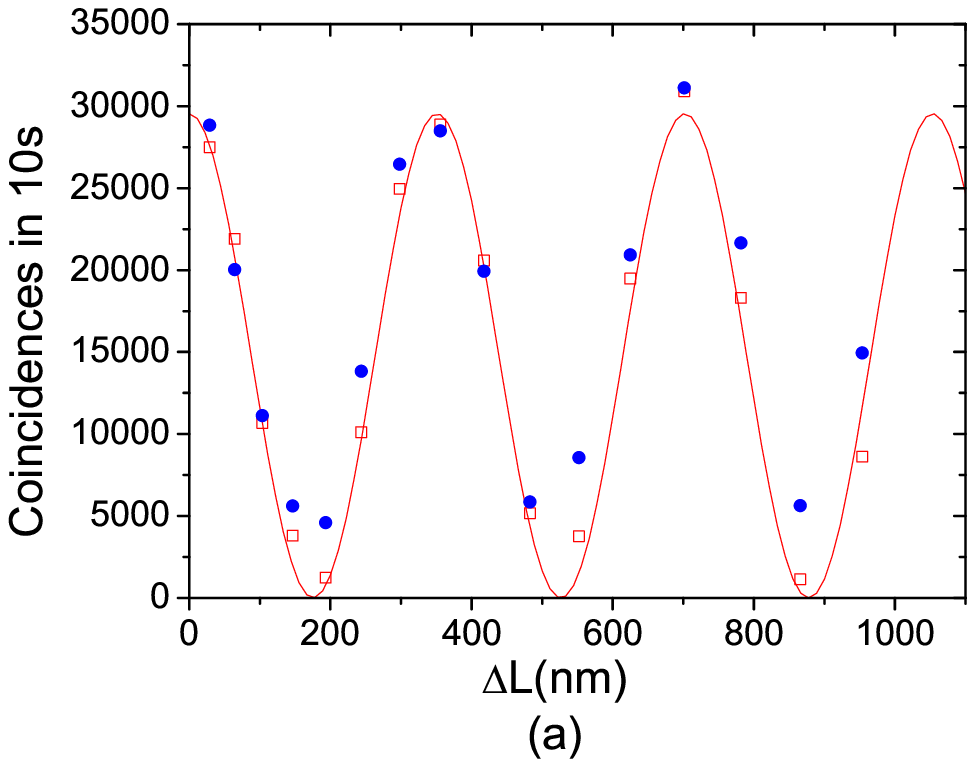}
\includegraphics[width=4.0cm,height=5.2cm]{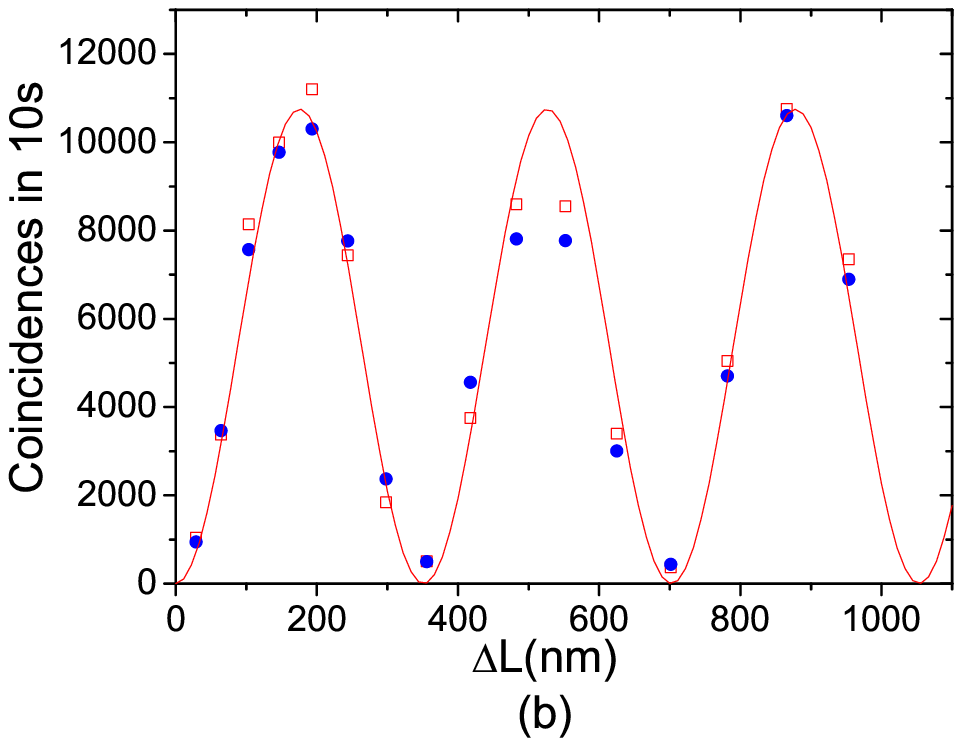}
\caption{(a) and (b) show the coincidences in the HV basis and HH
basis after the PBS respectively. The metal plate is removed from
the twin-lenses. Case 1(square dots), the first HWP is placed
before the twin-lenses. The solid line comes from a fit of the
experimental data with $R_{HV(VH)}=a /4(1+\cos (2\Delta \phi))$
using least square method. Case 2(round dots), the first HWP is
placed after the twin-lenses. The solid line comes from a fit of
the experimental data with $R_{HV,VH}=a /8(1-\cos (2\Delta \phi))$
using least square method. It can be seen that twin lenses without
metal plate has little effect to the interference patterns.}
\end{figure}

\begin{figure}[b]
\includegraphics[width=4.0cm,height=5.1cm]{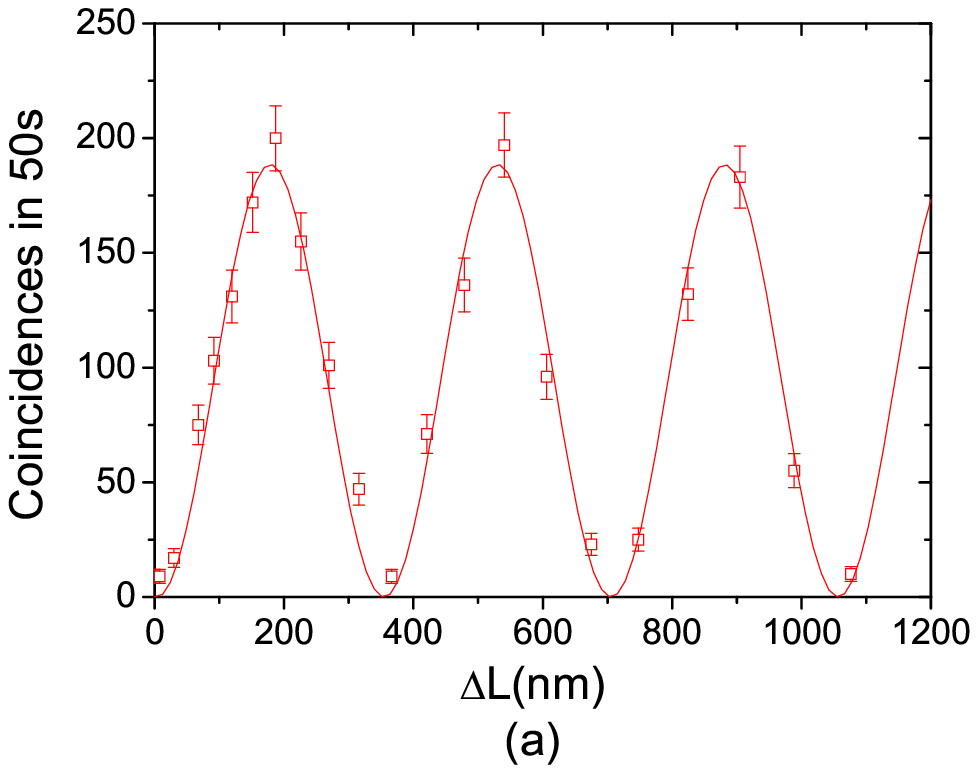}
\includegraphics[width=4.0cm,height=5cm]{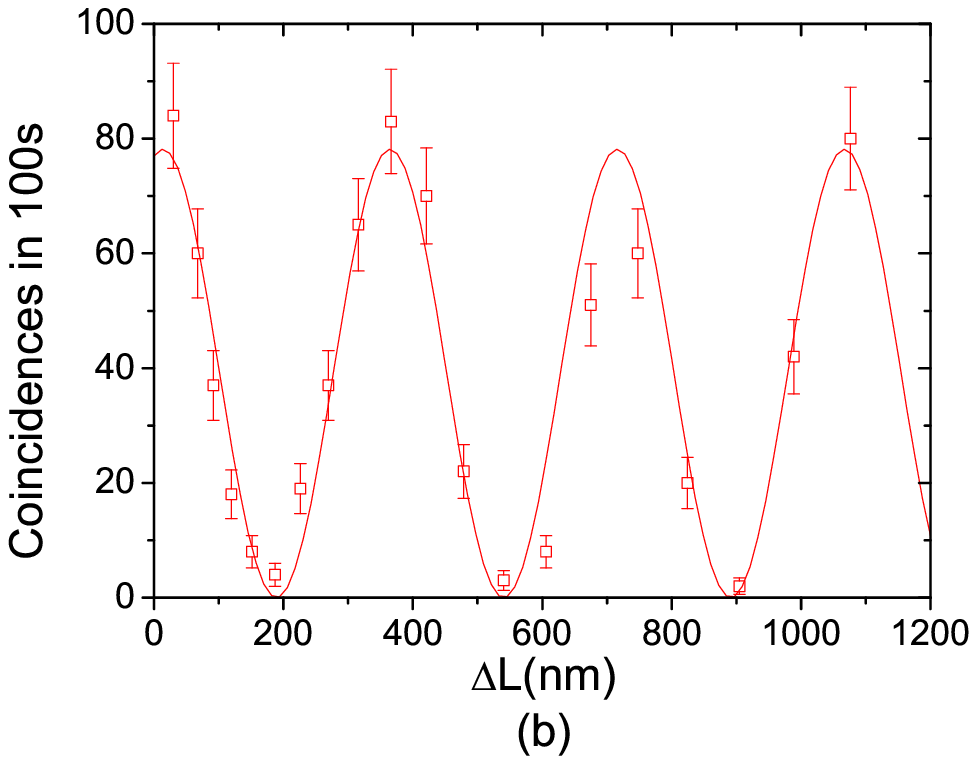}
\caption{(a) and (b) show the coincidences in the HV basis and HH basis
after the PBS respectively. The metal plate is placed between the twin
lenses, and the first HWP is placed before the twin-lenses, so the photon
pair acts on the hole array is in HH-VV polarization.}
\end{figure}

\begin{figure}[b]
\includegraphics[width=4.0cm,height=5cm]{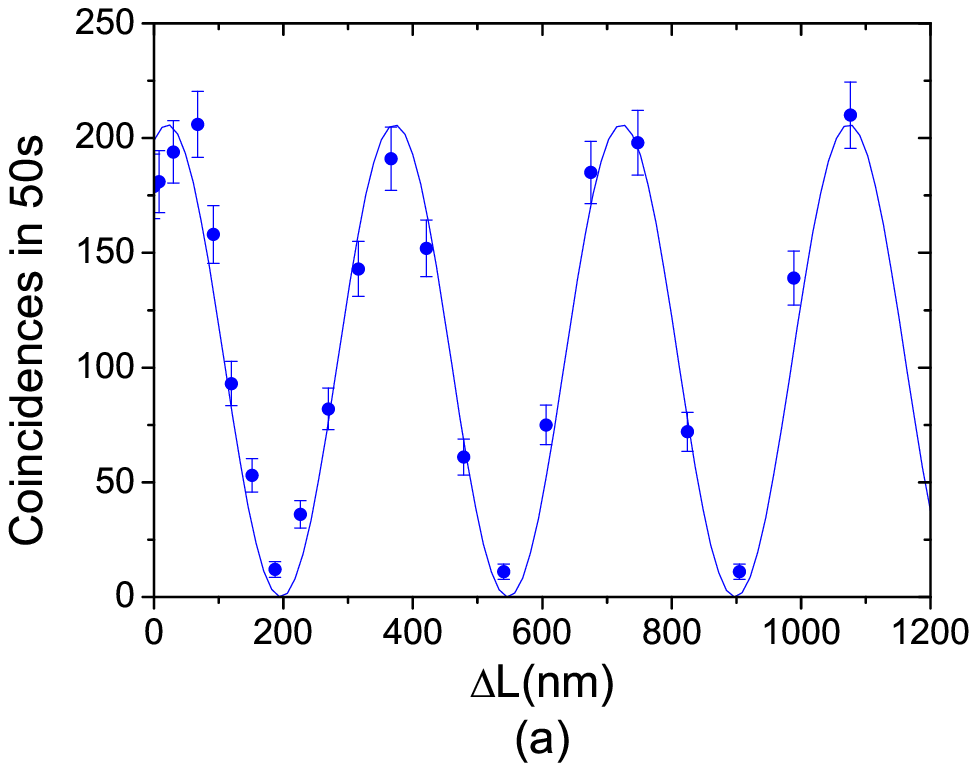}
\includegraphics[width=4.0cm,height=5.1cm]{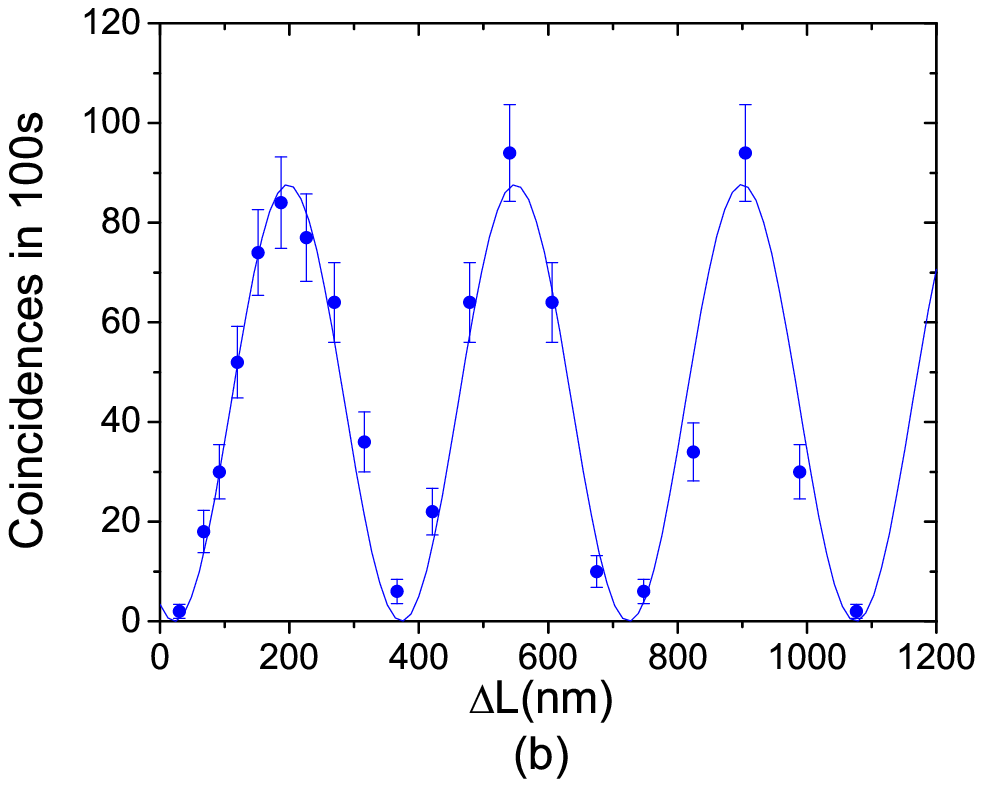}
\caption{(a) and (b) show the coincidences in the HV basis and HH
basis after the PBS respectively. The metal plate is placed
between the twin lenses, and the first HWP is placed behind the
twin-lenses, so the photon pair acts on the hole array is in HV
polarization.}
\end{figure}

Firstly, we consider the case when the metal plate is removed from
the twin-lenses. When the first HWP is placed either before or
after the twin-lenses, the two collinear down-converted photons in
the state $\left| HV\right\rangle $ will be changed into the state
$(\left| HH\right\rangle -\left| VV\right\rangle )/\sqrt{2}$. Then
these two photons travel two birefringent crystals, where they get
a phase difference $\Delta \phi $ between horizontal and vertical
polarization modes(In the figures below, we use $\Delta \L
$=$\Delta \phi /2\pi*702nm$ as axis label). The two-photon state
is thus in the form of $(\left| HH\right\rangle -e^{i2\Delta \phi
}\left| VV\right\rangle )/\sqrt{2}$. After passing the second
HWP(at 702nm), their state are transformed into $\frac
1{2\sqrt{2}}\{(1-e^{i2\Delta \phi })(\left| HH\right\rangle
+\left| VV\right\rangle )+(1+e^{i2\Delta \phi })(\left|
HV\right\rangle +\left| VH\right\rangle )\}$. Then the two photons
are separated by a polarization beam splitter (PBS). The
coincidence click of the two detectors directly behind PBS
projects these photons in the basis of different polarization
($\left| HV\right\rangle $ or $\left| VH\right\rangle $ due to the
indistinguishability of the two photons). The experiment result is
shown in Fig. 3a, which fits nicely with the theoretical
interference pattern of the coincidence rates in the $\left|
HV\right\rangle $ and $\left| VH\right\rangle $ basis
\begin{equation}
R_{HV(VH)}=\frac 14(1+\cos (2\Delta \phi)).
\end{equation}
In addition, we can place a lossless 50\%-50\% beam splitter with
two detectors behind one output port of the PBS and then project
photons in the $ \left| HH\right\rangle $ basis. The experiment
results is shown in Fig. 3b, which also fits well the theoretical
interference patterns of the coincidence rates
\begin{equation}
R_{HH}=\frac 18(1-\cos (2\Delta \phi)).
\end{equation}
The visibility of these interference patterns is about 93$\%$.
These results show that this Mach-Zehnder (MZ) type interferometer
can directly measure the de Broglie wavelength of collinear
polarization photons. The de Broglie wavelength of the
down-converted collinear biphoton is $\lambda /2$, agreeing with
the Fock state $\left| 2\right\rangle $ result. It can also be
seen that two lenses without metal plate has little effect to the
interference patterns.

Now we put the metal plate with hole array between the twin
lenses. First of all, we study the case when the two collinear
photons of identical polarization simultaneously incident on the
hole array. In this case, the first HWP is placed before the two
lenses containing the metal plate and changes the two photons from
the state$\left| HV\right\rangle $ to the state $(\left|
HH\right\rangle -\left| VV\right\rangle )/\sqrt{2}$. Then these
two collinear photons in the state $(\left| HH\right\rangle
-\left| VV\right\rangle )/\sqrt{2}$ simultaneously arrive the hole
array and excite surface plasmon. Similar to the case without
metal plate, the zero order re-eradiated photons, namely, the
surface plasmon assisted transmission photons, pass through a twin
birefringent crystal assembly, and get a phase difference between
horizontal and vertical polarization modes, before travelling to
the second HWP and being detected. Fig. 4a shows the interference
pattern of the coincidence rate between the two detectors directly
behind PBS projects. This corresponds to project them in the
$\left| HV\right\rangle $ or $\left| VH\right\rangle $ basis. The
resultant interference pattern is very similar to the case that
metal plate is removed, and can be nicely fitted to
\begin{equation}
R_{HV(VH)}=\frac 14(1+\cos (2\Delta \phi +0.98\pi ))
\end{equation}
Similarly, we also place a lossless 50\%-50\% beam splitter behind
one output port of the PBS and project these photons in $\left|
HH\right\rangle $ basis. The interference pattern is shown in Fig.
4b, which corresponds to

\begin{equation}
R_{HH}=\frac 18(1-\cos (2\Delta \phi +0.93\pi ))
\end{equation}
The visibility of these two interference patterns is about 92$\%$.
We can conclude the de Broglie wavelength of re-eradiated photons
in this case is $\lambda /2$.  Two photons are re-eradiated by the
surface plasmon, excited simultaneously by two collinear photons
of identical polarization.
The quantum coherence of the biphoton is preserved in this biphoton$%
\rightarrow $ surface plasmon $\rightarrow $biphoton conversion.

Secondly, we study the case when the two photons simultaneously
exciting surface plasmons have different polarization, by placing
the first HWP behind the metal plate. The down-converted collinear
photon pair in $\left| HV\right\rangle $ modes are directly
incident on the hole array. And the re-eradiated photons by the
surface plasmons, simultaneously excited by two photons of
orthogonal polarization, travel through the Mach-Zehnder (MZ) type
interferometer of collinear polarization photons so that their de
Broglie wavelength is measured. The resultant interference
patterns of projection in $\left| HV\right\rangle $ and $\left|
HH\right\rangle $ basis are respectively shown in Fig. 5a and Fig.
5b, which can also be nicely fitted respectively with
\begin{eqnarray}
R_{HV(VH)} &=&\frac 14(1+\cos (2\Delta \phi -0.12\pi )) \\
R_{HH} &=&\frac 18(1-\cos (2\Delta \phi -0.13\pi )).
\end{eqnarray}
The visibility is also about 92$\%$. The de Broglie wavelength of
re-eradiated photons is still equal to that of Fock state $\left|
2\right\rangle $. There are still two photons re-eradiated by the
surface plasmon, excited simultaneously by two collinear photons
of orthogonal polarization. The wave packets of the two photons
are kept superposed coherently in the transmission of biphoton in
$\left| HV\right\rangle $ mode.

If there is any decoherence in the plasmon-assisted transmission,
it corresponds to adding a random phase to the transmitted
photons. In the Ref.(1), two photons in $(\left| HV\right\rangle
-\left| VH\right\rangle )/\sqrt{2}$ mode are non-collinear. When
only one photon travelling through the hole array, it may add a
phase $\Delta \phi _H(t)$ ( or $\Delta \phi _V(t)$) to the single
photon in$\left| H\right\rangle $ ( or $\left| V\right\rangle $ )
mode, the experiment result of entanglement preservation shows
that $\Delta \phi _H(t)-\Delta \phi _V(t)=constant$ . In the
present experiment, two collinear photons simultaneously incident
on the hole array and excite surface plasmon. When the incident
two photons are in $(\left| HH\right\rangle -\left|
VV\right\rangle )/\sqrt{2}$ mode, the decoherence corresponds to
adding a phase of $\Delta \phi _{HH}(t)$ ( or $\Delta \phi
_{VV}(t)$) to the biphoton in$\left| HH\right\rangle $ ( or
$\left| VV\right\rangle $ ) mode. Due to the possible interaction
between the surface plasmons excited two photons, there is no
clear relationship between $\Delta \phi _{HH}(t)$ ( or $\Delta \phi _{VV}(t)$%
) and $\Delta \phi _H(t)$ ( or $\Delta \phi _V(t)$).  The present
de Broglie wavelength measurement shows that $\Delta \phi
_{HH}(t)-$ $\Delta \phi _{VV}(t)$ is a constant value. To
determine the exact value of $\Delta \phi _{HH}(t)$ and $\Delta
\phi _{VV}(t)$ (or $\Delta \phi _H(t)$ and $\Delta \phi _V(t)$), a
Mach-Zehnder (MZ) interference measurement of biphoton (or single
photon) with separate paths is needed. However, it is still
impossible to determine the random phase $\Delta \phi _{HV}(t)$ in
the decoherent transmission of biphoton in $\left| HV\right\rangle
$ even the MZ interferometer with separate paths.

We note that there is a $\pi $-phase (1.10$\pi $ and 1.06$\pi $
for HV coincidence and HH coincidence respectively) between the
case when the two photons is first transformed into the state
$(\left| HH\right\rangle -\left| VV\right\rangle )/\sqrt{2}$ by
the first HWP before incident on the metal plate, and the case
when the two photons in the state $\left| HV\right\rangle $ first
excite the hole array before travelling to the first HWP. Although
the detailed mechanism for it is not definitely identified, we
speculate that it is not the decoherence of hole array, but may be
accounted by the fact that our metal plate is actually made up of
four hole arrays of $15\mu m\times 15\mu m$ area and there is a
mismatch and in-homogeneity among different hole arrays. Thus this
metal plate has an effect similar to $\lambda /4$ birefringent
plate and can cause $\pi /2$ phase difference between horizontal
and vertical polarization mode. When the two photons in the state
$(\left| HH\right\rangle -\left| VV\right\rangle )/\sqrt{2}$
travel the metal plate, this birefringent effect add a phase $\pi
$ between $\left| HH\right\rangle $ and $\left| VV\right\rangle $
modes so that the re-eradiated two photons will be in the state
$(\left| HH\right\rangle -e^{i\pi }\left| VV\right\rangle
)/\sqrt{2}$. For the case that two photons in $\left|
HV\right\rangle $ mode pass the metal plate, the re-eradiated
photons will be in $e^{i\pi /2}\left| HV\right\rangle $. This
global phase has no effect to the de Broglie wavelength
measurement. Actually, we have checked this conjecture by the
single photon polarization state tomography measurement, which
shows the $\left| H\right\rangle +\left| V\right\rangle $ (or
$\left| H\right\rangle -\left| V\right\rangle $ ) mode photon
excited plasmon will re-eradiate photon in $\left| H\right\rangle
-i\left| V\right\rangle $ (or $\left| H\right\rangle +i\left|
V\right\rangle $ ) mode. Surprisingly, such birefringent effect of
hole array is not mentioned in the previous experiments. In
another metal plate with hole array fabricated in the same FIB
procedure, phase birefringent effect of about $\pi $ is observed.
Much more work is still needed to fully understand the
birefringent effect of the metal plate with medley hole
array\cite{guo}.

In conclusion, we have measured the de Broglie wavelength of
re-eradiated photons of the surface plasmons simultaneously
excited by two collinear photons of either identical or orthogonal
polarization, which shows that two photons are re-eradiated and
the quantum coherence of the biphoton is preserved in the
conversion process of transforming biphoton to plasmon and then
back to biphoton. The decoherence in two $\left| HH\right\rangle $
and $\left| VV\right\rangle $ mode photons transmission, if there
is any, cause similar random phases ( $\Delta \phi _{HH}(t)-\Delta
\phi _{VV}(t)=constant$ ). These results may give us more hints to
the understanding of the plasmon-assisted transmission of photons.

In the preparation of the present manuscript, we note that there
is an experiment showing quantum superposition and entanglement of
mesoscopic plasmons excited by time-bin entangled
photons\cite{ginis}.

\begin{center}
\textbf{Acknowledgments}
\end{center}

This work was funded by the National Fundamental Research Program
(2001CB309300), National Nature Science Foundation of China
(10304017), the Innovation Funds from Chinese Academy of Sciences.
ZYO is also supported by the US National Science Foundation under
Grant No.0245021.

\end{document}